%% file: 00-main.tex
\documentclass{article}

\usepackage{arxiv}

\usepackage{balance}
\usepackage[table,dvipsnames]{xcolor}
\usepackage{parskip}

\usepackage{graphics}
\usepackage{subfigure}
\usepackage{comment}
\usepackage{xparse}
\usepackage{wrapfig}

\newcommand{\text}{\mathrm}
\usepackage{amssymb,amsmath}
\usepackage{xparse}
\usepackage{color}
\NewDocumentCommand{\heng}{ mO{} }{\textcolor{red}{\textsuperscript{\textit{Heng}}\textsf{\textbf{\small[#1]}}}}
\usepackage[utf8]{inputenc} 
\usepackage[T1]{fontenc}    
\usepackage{hyperref}       
\usepackage{url}            
\usepackage{booktabs}       
\usepackage{amsmath}
\usepackage{amssymb}
\usepackage{amsfonts}       
\usepackage{nicefrac}       
\usepackage{microtype}      
\usepackage{graphicx}
\usepackage{caption}

\begin{document}

\title{The Paradox of Information Access: Growing Isolation in the Age of Sharing}


\author{Tarek Abdelzaher, Heng Ji, Jinyang Li, Chaoqi Yang,\\
Department of Computer Science,\\
University of Illinois at Urbana Champaign\\
\AND
John Dellaverson, Lixia Zhang,\\
Department of Computer Science,\\ 
University of California, Los Angeles\\
\AND
Chao Xu,\\
Department of Physics,\\ 
University of California, San Diego\\
\AND
Boleslaw K. Szymanski,\\
Department of Computer Science,\\
Rensselaer Polytechnic Institute}


%

\date{}

\maketitle


\begin{abstract}
Modern online media, such as Twitter, Instagram, and YouTube, enable anyone to become an information producer and to offer online content for potentially global consumption. 
By increasing the amount of globally accessible real-time information, today's ubiquitous producers contribute to a world, where an individual consumes vanishingly smaller fractions of all produced content.  In general, consumers preferentially select information that closely matches their individual views and values. The bias inherent in such selection is further magnified by today's information curation services that maximize user engagement (and thus service revenue) by filtering new content in accordance with observed consumer preferences. Consequently, individuals get exposed to increasingly narrower bands of the ideology spectrum. Societies get fragmented into increasingly ideologically isolated enclaves. These enclaves (or echo-chambers) then become vulnerable to misinformation spread, which in turn further magnifies polarization and bias. We call
this dynamic the {\em paradox of information access\/};\footnote{The authors first used the term {\em paradox of information access\/} in~\cite{xu2020paradox} (https://arxiv.org/abs/2004.01106), where they developed an analytic model of the paradox. While the current document shares introductory material with~\cite{xu2020paradox}, its main focus is on proposed mitigation strategies as opposed to the analytic model.} a growing ideological fragmentation in the age of sharing.
This article describes the technical, economic, and socio-cognitive contributors to this paradox, and explores research directions towards its mitigation.    

\end{abstract}

\input{01-IntroBackground}
\input{02-LixiaNDN}

\input{03-HengMisinfomation}

\input{031-HengConsistency}

\input{032-TarekBias}

\input{04-HengGeneration}

\input{05-Dialogue}

\section{Ethics, Incentives, and Regulation}
\noindent
Technological advances suggest that information distillation tools, powered by security, machine intelligence algorithms, and an understanding of human biases and morality, may be developed to export reliable provenance, distill opposing opinions, de-bias content, detect underlying moral underpinnings, and eventually summarize the arguments used according to different moral perspectives. Several questions remain. 

First, would such tools be ethical? On this issue, two viewpoints commonly arise. The first viewpoints observes the increased global emphasis on upholding user privacy. Distilling moral values from published content, for example, seems intrusive from a privacy perspective. In contrast, the second opinion notes that social media serve another key function, besides social access. Namely, they give individuals a voice. The prerogative to be heard is a key human right.  Thus, tools discussed above can, in fact, reach across echo-chambers, empowering individuals to reach larger populations. 
This observation leads to the next question. What would incentivize information consumers to use such tools? 
Can a better understanding of human cognitive biases help develop a more persuasive technology that not only addresses the concerns discussed above, but also offers better value to the user than the current alternatives? This remains an open design challenge at the present time. 

A related question is whether it might be appropriate to regulate the design of information distillation technologies to ensure that they have certain desirable and measurable properties. Many infrastructure technologies that affect large populations are regulated using appropriate legal frameworks. For example, water purification infrastructure must generally meet certain minimum filtering quality standards. Traffic management on the public Internet (in the US) was subjected to net neutrality requirements until 2018 (meaning that providers could not discriminate in favor of one network traffic category over another). Yet, information curation services that filter or expose large volumes of information to inform users are not presently bound by any filtering quality or neutrality requirements. This is possibly, in part, due to the lack of appropriate metrics to quantify content quality and neutrality, and the lack of appropriate technological means to measure such metrics. Development of measurable metrics might be another research challenge.     

\section{Conclusions}
\noindent
Concerns with growing polarization and fragmentation have been reported in literature for a long time. Several factors contributing to polarization are already well-understood, such as homophilly and network topology. This article presented another grim reality: the {\em mere volume of generated information\/} (especially with small infusions of misinformation) significantly exacerbates polarization and ideological fragmentation.
Analysis of this phenomenon suggested several mitigation strategies, including mechanisms to ensure strong/secure content provenance, algorithmic fact-checking solutions to remove misinformation, techniques that mitigate curation bias, and overload mitigation strategies, such as abstractive summarization.
Modern technical advances offer capabilities that can help construct such tools and formulate relevant optimization problems. Issues related to incentives, ethics, and technology persuasion remain to be resolved. 
The work suggests the need for continued research to stop the described paradox: disentangle increases in information volume and access from unintended consequences on ideological fragmentation. 

\section*{Acknowledgment}
\noindent
The work was funded in part by DARPA under contracts W911NF-17-C-0099 and FA8750-19-2-1004, and DTRA under contract HDTRA1-18-1-0026.



%
\balance

\bibliographystyle{IEEEtran}
\bibliography{paper,ref-heng}

\end{document}

%% file: 01-IntroBackground.tex


\section{Introduction}
\noindent
In human psychology, several well-known paradoxes exist when monotonicity of reward with respect to a beneficial stimulus is ultimately broken. For example, {\em the paradox of choice\/} maintains that proliferation of choices ultimately leads to decreased satisfaction, as individuals perceive a higher opportunity cost in committing to their decisions~\cite{schwartz2004paradox}.
We argue that modern online social media platforms that allow publishing information for potentially global access create a different type of paradox. Namely, the {\em mere availability of access to more information ultimately increases fragmentation of society into ideologically isolated enclaves\/}. We call it the {\em paradox of information access\/}. 

\begin{figure*}[t]
    \centering
    \includegraphics[width=0.9\linewidth]{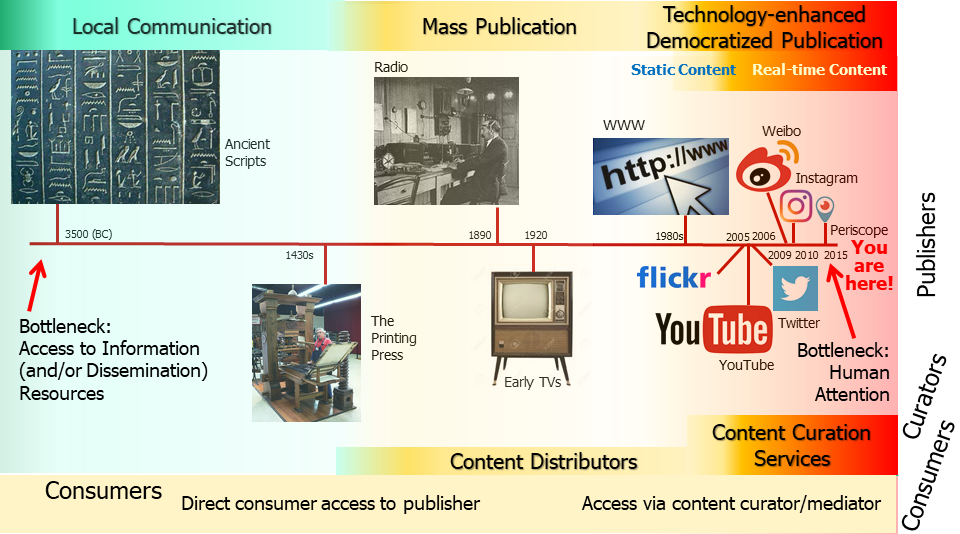}
    \caption{Evolution of Information Flow: The emergence of information overload due to ubiquitous access, and the rise of information curation services that mediate information visibility.}
    \label{fig:flow}
\end{figure*}

This paradox is brought about by the disruptive change in content dissemination dynamics that occurred over a relatively short period of time in human history. 
Much in the way the invention of the press five hundred years ago revolutionized information dissemination, the advent of modern online social media platforms, as shown in Figure~\ref{fig:flow}, provides everyone with ample opportunities to publish near real-time information. We will refer to this development as {\em democratization\/} of publishing and access.  
Public information available on social media platforms is intended to have {\em global visibility\/}.
Yet, the resulting increased volume of information (created by a vastly enlarged number of diverse producers) makes {\em human attention\/} an increasingly over-subscribed bottleneck. It necessitates information filtering, as recipients must make consumption choices. Information filtering is aided by algorithmic information curation services (e.g., search engines, ranking engines, and information recommendation systems) whose business model maximizes recipient engagement (e.g., clicks). 
In effect, information curation services optimize a self-loop; they give consumers more of what these consumers express interest in. When applied to information flow, this optimization leads to a situation where our existing biases impact how future information is prefiltered for our consumption, leading to feedback that reinforces these biases. The very channel that makes information sharing ubiquitous therefore mediates information visibility (e.g., by information ranking), preferentially offering different populations different content. This bias-reinforcing filtering, tailored to consumer biases, gradually erodes the common ground for dialogue among communities of different beliefs, creating the paradox: {\em ideological fragmentation in an age of democratized global access\/}.

\begin{figure*}[htb]
    \centering
    \includegraphics[width=0.8\linewidth]{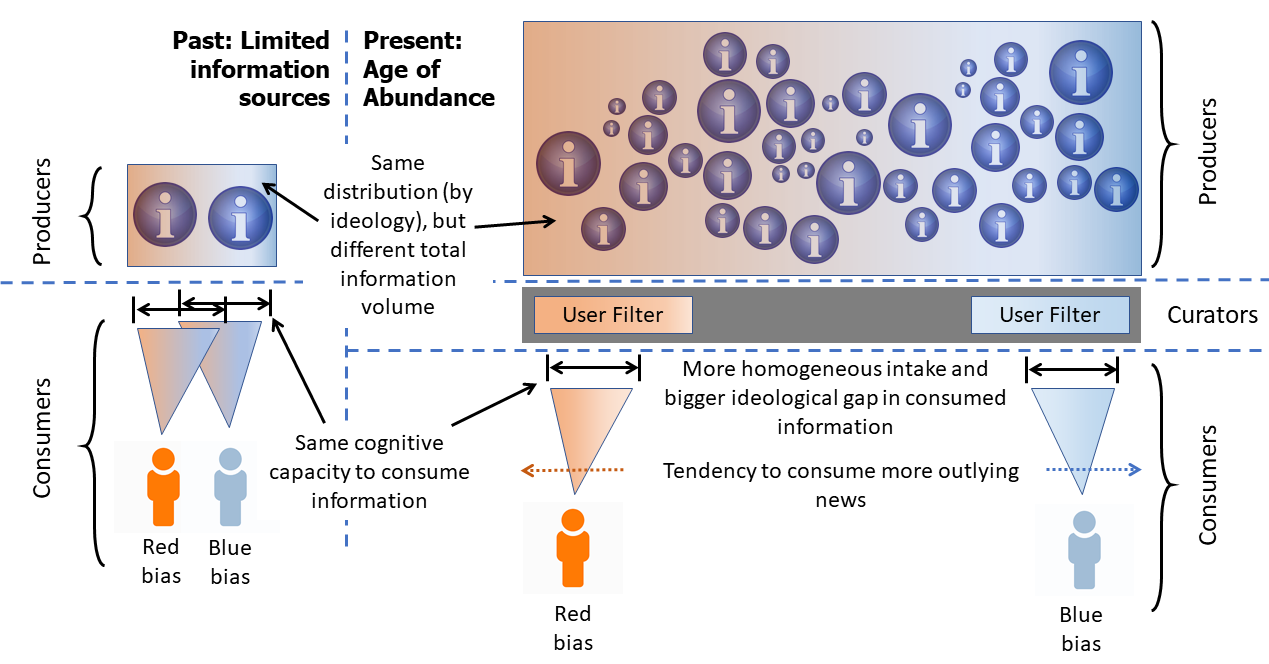}
    \caption{Notional illustration, from~\cite{xu2020paradox}, demonstrating impact of information volume on the evolution of bias. Assuming that information curation services will recommend information that accurately matches consumer bias, given the same amount of consumed information, content given to a particular consumer will be much more homogeneous (on the ideology spectrum), and content given to different consumers may have bigger ideological gaps, compared to a situation with more limited sources.}
    \label{fig:gap}
\end{figure*}

Figure~\ref{fig:gap} notionally demonstrates the effect of increased information volume. As the volume of accessible information increases, the fraction of it that an individual consumes (i.e., our individual {\em coverage\/}) decreases. If consumers choose to access information that is closer to their beliefs, the decreased coverage (centered around their belief) implies exposure to progressively narrower 
slices of the overall belief distribution. Larger ideological gaps emerge between information consumed by different parties. The narrowing exposure and growing ideological gaps facilitate the spread of disinformation, adding significant fuel to the polarization dynamics. 

Democratized access to information implies democratized access to {\em misinformation\/}. Thus, the channel that facilitates information sharing also empowers misinformation dissemination. A comprehensive solution to the problem therefore needs not only to address information bias but also suppress misinformation spread. 

The dynamics described above have significant societal consequences. In fact, one might go as far as saying that democracy itself, as conceived by modern society, is in danger. This danger might arise from actions of adversaries who may find it easier to ignite societal conflicts. More disturbingly, it may also arise internally from {\em selective exposure\/} to true facts (e.g., ``police officer shot youth,'' omitting that the youth shot at police officer first). 
Democracy in modern society is based on the assumption that constituents are {\em well-informed\/}. In an environment, where this assumption is violated due to disinformation or incomplete views, the foundations of democracy themselves may be jeopardized. 

\section{The Information Landscape}
\noindent

To appreciate the fundamental nature of the paradox, it is useful to explain it as an emergent behavior that arises (in the age of information overload) from interactions among three constituencies, driven by their own self-interest; namely, content {\em producers\/}, {\em curators\/}, and {\em consumers\/}. While the real world is significantly more complex, in order to develop a basic intuition into its key dynamics, we oversimplify it by the following hypotheses: 



\subsection*{The Production Hypothesis}

{\em Content producers aim to maximize the dissemination footprint of their content and, as such, are incentivized to promote broader dissemination to their target population.\/}

Competition for human attention creates incentives for producers to develop content that is more noticeable and attractive for consumption. It may encourage a form of content engineering that ranges from relatively benign (e.g., a focus on negative news because it propagates faster~\cite{fiske1980attention,shoemaker1996hardwired}) to outright malicious (e.g., faking information provenance or spreading intentional disinformation). The problem is further exacerbated by the lack of adequate source identity verification on current social media, allowing misattribution of  (dis)information. Manipulated content reduces the fraction of human attention spent on consuming other (more genuine) news~\cite{Sreenivasan2017Predictive}. The ramification is {\em reduced overall quality of consumed content.\/} 

\subsection*{The Curation Hypothesis}
{\em Content curators are incentivized to maximize client engagement with their curated content (e.g., clicks).\/}

Given the increased generated volume and diversity of information, content filtering plays a bigger role in shaping the overall information flow. Driven by business revenue, curators shape content visibility in a manner consistent with consumer preferences, thereby reinforcing consumer bias. The  ramification is {\em increased bias in information visibility.\/}

\subsection*{The Consumption Hypothesis}

{\em Content consumers aim to manage their own overload and, as such, are incentivized to engage in practices that reduce cognitive load, while satisfying their personal interests in content.\/}

Modern information propagation behaviors are a combination of ingrained biases and efficiency shortcuts, exacerbated by overload. For example, we increasingly seek (and spread) more sensational and surprising news~\cite{lamberson2018model,varshney2019must}. Seeking (and spreading) more outlying or surprising news is a collective {\em learning efficiency\/} tactic in that it avoids expending cognitive resources on processing redundant inputs. 
The tactic seems increasingly pronounced as the volume of information increases~\cite{varshney2019must}, arguably biasing our collective attention towards more extreme content. Imitation is another efficiency tactic~\cite{bonabeau2004perils}. Replicating behaviors of role models and like-minded individuals obviates expending one's own cognitive resources (e.g., on ascertaining veracity of information we forward). Evidence suggests that its usage becomes more pronounced under time pressure~\cite{buckert2017imitation}. We are thus more prone to replicating more outlying/extreme content and doing so with less verification, which {\em increases prevalence of more extreme (mis)information flow\/}.

\subsection*{The Rise of Information Disorder}
\noindent
To above discussion suggests that we suffer from increasingly narrower and ideologically more biased exposure ({\em see ramification of the curation hypothesis}) to lower-quality information ({\em see ramification of the production hypothesis}) without adequate verification ({\em see ramification of the consumption hypothesis}). These dynamics leave populations vulnerable to the emergence of disinformation that further breeds ideological biases and polarization. In turn, biases make it easier to believe further disinformation that is tailored for these biases. This detrimental cycle needs to be understood, and solutions need to be constructed that stop it.

\section{Analytical Modeling}
\begin{figure}[htb]
    \centering
    \includegraphics[width=0.5\textwidth]{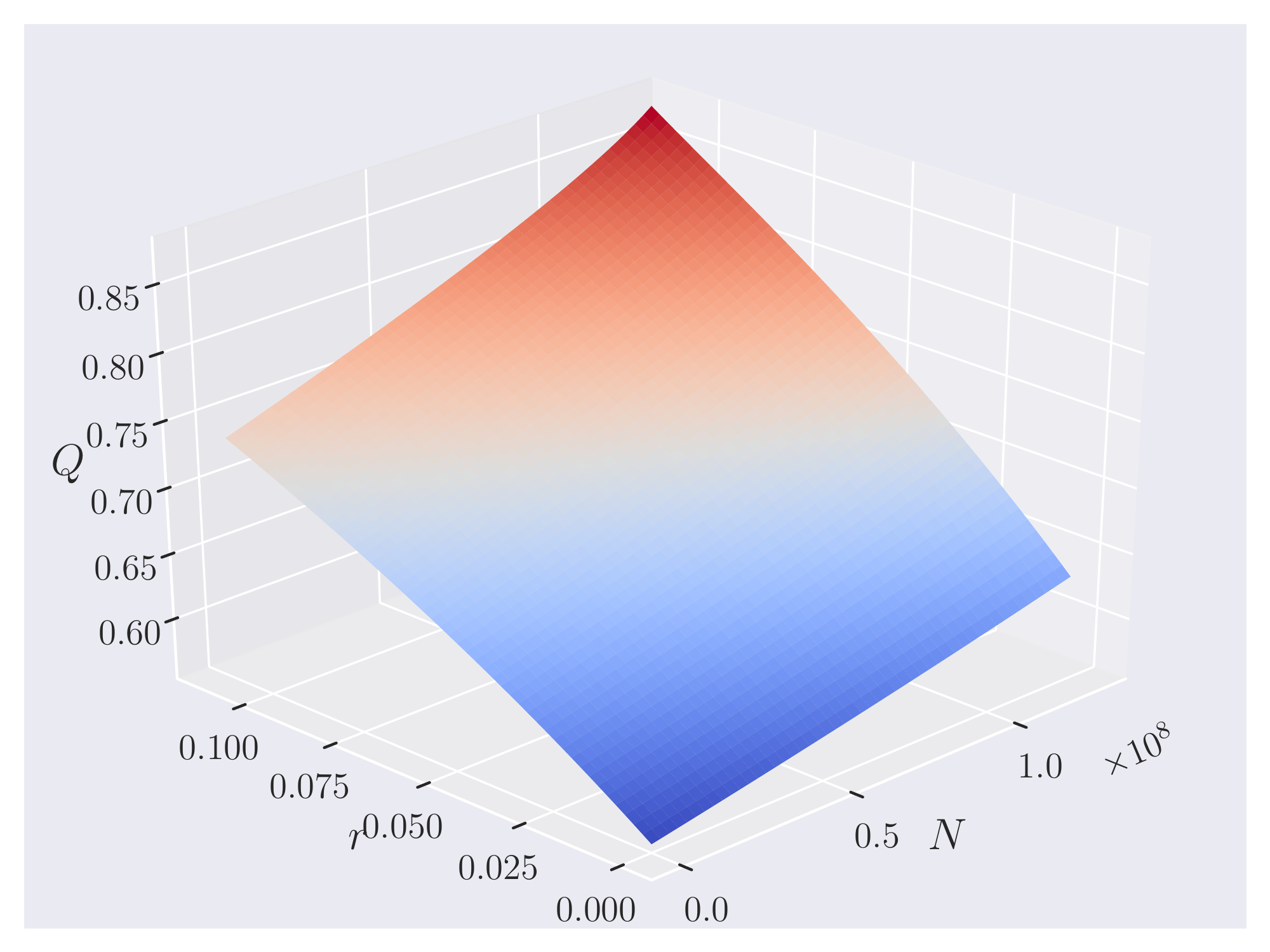}
    \caption{The figure, originally from~\cite{xu2020paradox}, plots population bifurcation (where higher $Q$ means more polarization) versus total information volume, $N$, and the ratio of injected misinformation, $r$. See~\cite{xu2020paradox} for other settings.}
    \label{fig:load}
\end{figure}

While significant literature exists on the factors that contribute to polarization (such as homophilly, bias, and network topology)  none explicitly focuses on the impact of information overload.
One exception is a recently developed model that studies the impact of information volume on the emergence of polarization~\cite{xu2020paradox}. The model, simulated in Figure~\ref{fig:load}, confirms the emergence of polarization with increased information overload. It also confirms that small amounts of plausible misinformation (outlying content that is not distinguished from genuine content), emitted by committed agents (those whose belief is not influenced by external forces) can significantly increase the degree of polarization.  

\begin{figure*}[htb]
    \centering
    \includegraphics[width=0.9\textwidth]{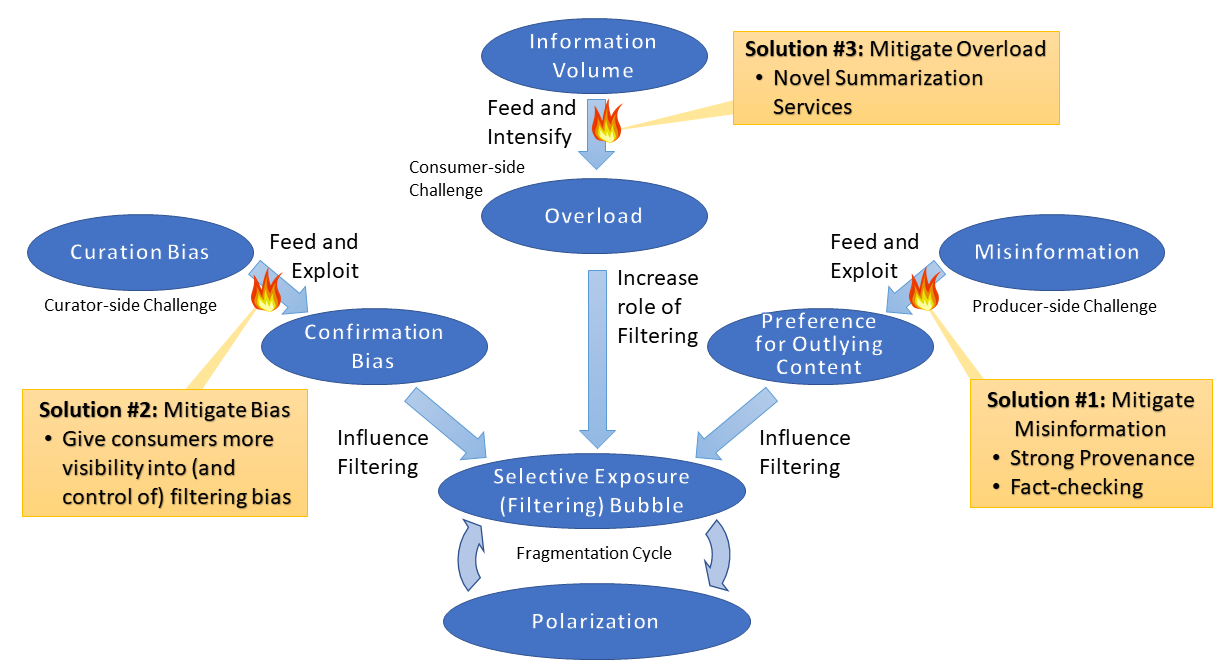}
    \caption{Polarization Influencers and Mitigation Mechanisms.}
    \label{fig:cycle}
\end{figure*}

The model represents consumers as particles in a belief space, whose beliefs (i.e., positions in the space) are influenced by a combination of forces derived from the content they choose to engage with. Democratization of the publishing medium means that consumers are also producers. Changes in their positions are thus reflected in the distribution of subsequently shared content. The contribution lies in developing force expressions that adequately model factors from social psychology, such as confirmation bias and social influence, that impact a node's motion in the belief space. This motion is shown to follow a particle diffusion-drift model that reduces to a nonlinear Fokker–Planck 
equation~\cite{risken1996fokker}, commonly used in the study of stochastic systems such as fluid dynamics. Solving for the steady state, the equation reveals a bifurcation of population density in the belief space (i.e., emergence of polarization), whose depth increases with information volume and misinformation. Figure~\ref{fig:load} plots this relation.
Next, we describe how to mitigate these effects.


\section{Possible Solutions}
While analytical models can help understand the emerging dynamics, to mitigate the cycle, actionable solutions are needed. Fundamentally, three forces are at play: (i) producer-side content manipulations (e.g., that cater to our preference for outlying content), (ii) curation algorithms (that reinforce consumer bias), and (iii) the sheer information volume (that magnifies filtering behaviors, such as those driven by the above preferences and biases, thereby intensifying their impact). This suggests a three-pronged solution space, as shown in Figure~\ref{fig:cycle}.
These approaches are:

\begin{itemize}
    \item {\em Mitigating producer-side misinformation:\/} Figure~\ref{fig:load} suggests that even small infusions of misinformation have a particularly detrimental effect. Solutions are needed that detect (or make it harder to insert) bots, fake sources, and disinformation. In other words, one must prevent malicious exploitation that feeds the fragmentation cycle. 
    \item
    {\em Reducing curation bias:\/} While we cannot change human psychology, we can mitigate bias-reinforcing effects of current content curation services by giving consumers more direct control over the content they ingest.
    \item {\em Improving consumer-side information coverage:\/} Figure~\ref{fig:load} suggests that fragmentation increases with overload (due to an individuals' narrower information coverage). Solutions are needed that fix the {\em low coverage\/} problem. Abstractive multimodal summarization algorithms might be a possible avenue to decrease the effective gap between  available information and consumers' cognitive capacity.
\end{itemize}

Figure~\ref{fig:cycle} shows the relation among the above mitigation strategies. One should note that these remedies are not necessarily intended to work as an integrated single mechanism. Rather, as shown in the figure, they attack different links that feed the  fragmentation cycle (much the way different medications and social distancing remedies might inhibit different pathways for virus replication). In the following sections, we discuss the above mitigation strategies, respectively.    We also conjecture on possible technological solutions that might help bridge existing ideological gaps to help improve mutual understanding. 

%% file: 02-LixiaNDN.tex
\section{Mitigating Misinformation}
Two schools of thoughts arise in fighting misinformation. One develops algorithmic techniques to detect false elements in communicated narratives. Another redesigns the underlying channel in the first-place. We first describe how information-centric networking can help significantly reduce malicious content. We then show how misinformation detection algorithms can be designed on top to distinguish good and bad sources and information items.

\subsection{Restoring Channel Integrity}
Supporters of channel-centric solutions argue that any attempt to ameliorate producer-side information disorders must start with securing data provenance; without knowing the source of a piece of novel information, there is no way to reason about whether it is factual. Similarly, without a way to guarantee that information hasn't been manipulated since it was first produced, trust becomes much more complicated. In traditional media, one could verify the provenance of a piece of information more easily; while reading a newspaper, it is generally obvious where a piece of information comes from -- the author is stated at the top of the article, and their identity is vouched for by the organization that runs the newspaper. 
In today's social media ecosystem, every single one of the billions of netizens has become a potential news source. 
Ascertaining trust in content has become more difficult.

A possible solution to this problem is proposed in the context of named data networking~\cite{zhang2014named} (NDN).
NDN is a Future Internet architecture (an alternative to TCP/IP) that takes the position that every piece of content ought to be both named and cryptographically verifiable: every piece of data is accompanied by a digital signature that proves the data's integrity and provenance. In particular, this contrasts with the predominant security paradigm, which involves securing a channel for communication, rather than data. 

In a world where we only secure channels, we can only tell whom we heard something from. In a world where we secure data, we can tell who originally created something \emph{and who we heard it from}. The tendency of misinformation to spread over social media illustrates the important difference between these two concepts. Spam email chains illustrate this problem nicely: they are forwarded on by friends (that is to say, people who are trusted), but contain content that wasn't produced by them, and which shouldn't be trusted. Strong provenance allows us to distinguish between these two properties easily and naturally \cite{NDNTech}. 



One obvious tradeoff in requiring data provenance is that it prevents strong anonymity. We believe that this design choice is nevertheless well-worth it. 
Bitcoin, which advertises itself as being a provably anonymous, secure currency, has demonstrated that anonymity can lend itself to serious malfeasance. 
In 2018 Foley, Karlson, and Putniņš found that 44 percent of Bitcoin transactions were associated with illegal activity \cite{bitcoinCrime}. 
Anonymity can instead be offered as a service through authenticable anonymity providers, so that when an anonymous user violates the law, there are effective means to to capture the violators and keep society in order.




%% file: 03-HengMisinfomation.tex
\subsection{Algorithmic Misinformation Detection}
Securing provenance does not automatically remove misinformation on the channel. It only allows creating identities that can earn trust over time. It remains to reason about accuracy of information originating from different sources. 

While a plethora of solutions have been proposed that learn feature sets correlated with misinformation, the fundamental solution to detecting misinformation (besides relying on reputation of sources or incidental features of content) is to detect inconsistencies. In a world where knowledge elements are highly interdependent and correlated, it should, in principle, get progressively harder to insert a misinformation stream that is entirely consistent with all prior knowledge. Thus, one may imagine future algorithms that rank content not only based on reputation-weighted support (e.g., Google PageRank), but also based on the degree to which it is {\em mutually consistent\/}. While such algorithms may consider coarse-grained elements, such as web pages or sources, as features to help rank trustworthiness, the new framework will also assign separate trustworthiness scores to much finer-grained {\em knowledge elements\/}. 

Recent advances in natural language processing allow entity discovery and linking for multiple data modalities and languages.  
It is thus, possible to leverage complementary information from multiple modalities to expand, align, and ground knowledge elements and ensure consistency of published information.   
Recent work has developed multi-modal knowledge graph extraction methods~\cite{LiZareian2019} that can learn a unified representation for event triggers and arguments for both text and image modalities. 
The work demonstrated significant performance gains in knowledge  extraction, compared to the state-of-the-art single-modality solutions or multi-modal methods that use flat representations without graph structures.


%% file: 031-HengConsistency.tex
One may distinguish three levels at which inconsistencies can be detected in multimodal content analysis: (i) inconsistency internal to each knowledge element, (ii) inconsistency with background knowledge, and (iii) inconsistency across elements.

\begin{figure}[!hbt]
    \centering
  \includegraphics[width=\linewidth]{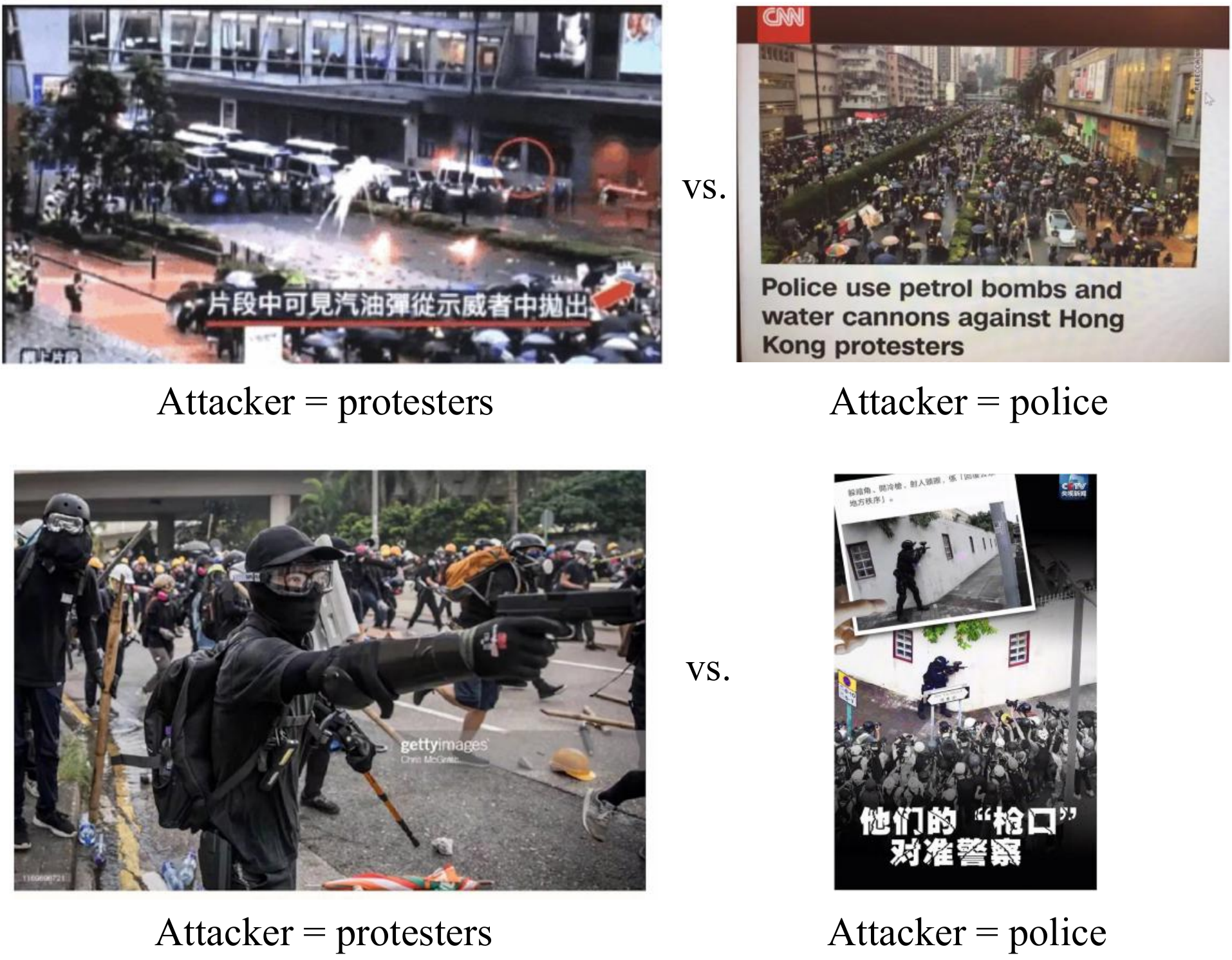}
  \caption{Cross-media Consistency Reasoning}
  \label{fig:crossmediaconsistency2}
\end{figure}

{\bf Internal Consistency Reasoning:\/}
Prior work~\cite{Hanselowski2018} found that simple textual measures such as readability and sentiment are effective at detecting social media posts spreading false information and rumors on document-level. Beyond such surface textual features, one can measure discourse coherence based on linguistic and discourse representations~\cite{McNamara2004}. 
Compared to images, disinformation detection from text at the single-asset knowledge element level is more challenging, especially when the text is written by a generally trustworthy source. 
For example, CNN published a news article titled ``Police use petrol bombs and water cannons against Hong Kong protesters". However, the original video (Figure~\ref{fig:crossmediaconsistency2}) revealed protesters threw bombs at the police, so CNN has apologized for its ``erroneous" reporting. This example demonstrates that by linking knowledge elements from multiple modalities, and automatically assigning a trust score to each element, one may detect inconsistent information.  

{\bf Background Consistency Reasoning:}
One may further compare multimodal knowledge elements against background knowledge (e.\,g., entity profiles, event temporal attributes, schemas and evolving patterns) extracted from historical data. Our recent work has constructed comprehensive background event-centric knowledge graphs from 15 years of multimedia multilingual news articles~\cite{Li2019Cross}. Using such background, one may refine veracity analysis. For example, if we extract the list of instruments employed by police in protest events, we can determine whether a specific instrument allegedly used according to some report (e.g., a slingshot in the alleged case of a recent Hong Kong protest) is likely. Furthermore, if we know the size of an area (e.g., Victoria Park), we can determine the range of the number of protesters who may gather there. Consistency of extracted information elements with such background knowledge may be instrumental in identifying false information.

\begin{figure}[!hbt]
\vspace{-1em}
\centering%
\includegraphics[width=\linewidth]{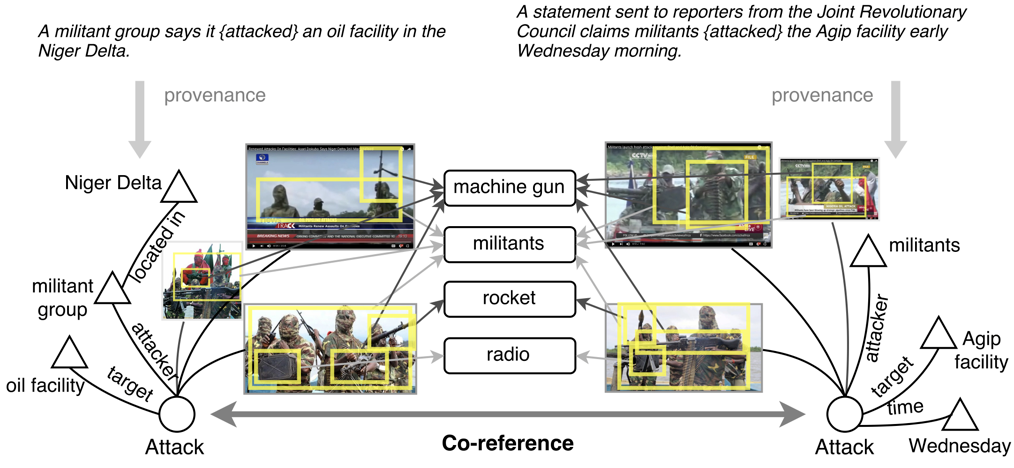}%
\vspace{-10pt}
\caption{Cross-document KG-based consistency reasoning.}
\label{fig:crossmediaconsistency}
\vspace{-18pt}
\end{figure}

{\bf Intra-element Consistency Reasoning:}
For collections of multimodal knowledge elements, one can take full advantage of the rich semantic labels and structures in the knowledge graph for linking and clustering across elements, beyond surface textual and visual feature encoding. Figure~\ref{fig:crossmediaconsistency} illustrates how knowledge elements extracted from images and videos can help merge two coreferential events for which two state-of-the-art text-only entity and event coreference resolvers failed to build the link between these two events.

For example, if two coreferential events reported by different data modalities contain different sentiment (e.g., violent group vs. cheering) or different weather conditions (rainy vs. sunny), these two events can be flagged as an inconsistent pair. One notable challenge is to distinguish disinformation from an event whose nature changes over time (e.\,g., the Baltimore 2014 protest turned violent after Freddie Gray died).  
Unseen data may quickly invalidate static event coreference and linking measures including salience, similarity, and coherence. 
Techniques for assessing compatibility of assertions (from both sensors and unstructured sources) have been described in previously developed truth-finding frameworks \cite{Huang2012,Zhi2015}. 


%% file: 032-TarekBias.tex
\begin{table*}[htb]
\centering
{\footnotesize
  \renewcommand{\arraystretch}{1.3}
  \caption{Top 5 Tweets from the Separated Polarities (Eurovision)}
  \label{eurovision_showcase}
  \centering
  \begin{tabular}{|p{0.003in}|p{2.65in}|p{2.65in}|}
    \hline
    & \multicolumn{1}{c|}{\bfseries{Pro-Jamala}} & \multicolumn{1}{c|}{\bfseries{Anti-Jamala}} \\\hline
    1 &
    Incredible performance by \#Jamala, giving Crimean Tatars, suffering persecution \& abuse, reason to celebrate https://t.co/XWOZADrywH
    &
    Eurovision rules stipulate a song must be NEW (since Sept). @Jamala sang “1944” in May 2015 https://t.co/T8Pu21hPiM https://t.co/rwAFHHIllf
    \\ \hline
    2 &
    Breaking: \#Russia launches harassment campaign against \#Jamala's @Twitter a/c. All known Kremlin trolls. @BBC\_ua @AP https://t.co/btk9QyUpkH
    &
    Eliot, @EliotHiggins, Jamala openly performed that song at a concert in May 2015, it was published online. It's a rule violation \#Eurovision
    \\ \hline
    3 &
    President awarded @jamala title of the People's Artist of Ukraine https://t.co/2df8J9zHP5
    &
    \#BOOM Jamala released Eurovision song commercially on 19.06.2015 in Kiev club Atlas. @EBU\_HQ https://t.co/LGbgb77RzH https://t.co/9Se1rwEkIg
    \\ \hline
    4 &
    Jamala's father: We do not talk with Russian journalists https://t.co/EEewVpZ9D2 https://t.co/51rcf8Je3K
    &
    \#Oops Poroshenko accidently confirms on TV that Jamala's Eurovision song 1944 is the same song ``Crimea is ours'' from May 2015. @EBU\_HQ
    \\ \hline
    5 &
    Congratulations to Ukraine on winning \#Eurovision 2016! @JAMALA wrote and composed her song `1944' by herself. https://t.co/vZjYHvtoC
    &
    \#Ukraine's singer, Jamala, has performed her \#Eurovision song a year ago, which is normally against the rules https://t.co/bxgeOK6X8D
    \\ \hline

  \end{tabular}
}
\end{table*}

\section{Reducing Curation Bias}
Discovering bad sources and misinformation, as described above, is not sufficient to break the paradox of information access. It merely prevents exploitation of consumer susceptibility to information disorders. A different solution is needed to undo reinforcement of consumer bias that plays a key role in promoting the paradox. 

One solution to eliminating curation bias is to put control over content visibility back in the hands of the consumer. Sir Tim Berners-Lee's Solid framework\footnote{https://solid.inrupt.com/} is one such attempt. Solid aims to give users agency over content they share, offering an entirely decentralized user-driven solution to the curation problem. In such a decentralized system, one can imagine content of different biases propagating within different groups. This observation leads to an interesting question: can one exploit content propagation patterns themselves to develop new services that help de-bias content, or at least give consumers visibility into (and control over) the degree of bias in the content they ingest? Research has shown, for example, that drivers adopt better driving habits when real-time feedback is given to them on the fuel consumption associated with their driving. Would exposing a metric that quantifies information bias (as well as knobs to control it) encourage changes in information consumers' behavior? 

Recent work has shown that it is possible to design algorithms that automatically separate different stances in a conflict based on the corresponding information propagation patterns~\cite{al2017unveiling}. In general, automatically separating different stances on a conflict is hard because it requires contextual knowledge. For example, when discussing  the Egyptian National Democratic Party, a statement against Mr. Kamal El-Ganzouri cannot be properly interpretted as supporting the party or not, unless Mr. El-Ganzouri's political affiliation is also known. Separating claims by their propagation patterns circumvents the need to understand local context. Groups of different biases propagate information differently, because each largely propagates items confirming their specific bias. It is therefore possible to construct a graph, where each individual is a node, and where individuals who share more similar content (e.g., propagate more of the same tweets) are ``closer'' together. Clustering the nodes in this graph separates the conflicted sides involved. Furthermore, identifying the most popular content propagated by each side determines the biases of that side. 

Table~\ref{eurovision_showcase} (originally published in~\cite{al2017unveiling}) 
presents sample results of applying the aforementioned algorithm to a Twitter data set collected after Eurovision 2016; a European song contest that took place in Stockholm, Sweden, in May 2016. It was won by a Ukrainian artist, Jamala, over a Russian runner-up, for a song with political (anti-Russian) undertones. This event is a good example of urban events that generate wide signatures on social media and, in this case, wide-spread controversy as well. 

The two columns in Table~\ref{eurovision_showcase} were separated automatically by observing information flow. 
The table demonstrates that clustering users and content by propagation patterns is very good at bias separation, despite the fact that the algorithm does not ``understand" the political context. It even correctly classifies bias in tweets that need situation-specific background to understand. For example, consider the tweet: {\footnotesize {\tt "\#BOOM Jamala released Eurovision song commercially on 19.06.2015 in Kiev club Atlas. @EBU\_HQ https://t.co/LGbgb77RzH https://t.co/9Se1rwEkIg"\/}}.  Unless one knows that Eurovision rules require that all song entries be {\em original\/}, it is hard to tell that the tweet is anti-Jamala. (It implies that her song should be disqualified because it appeared in another forum earlier.) 
Following propagation patterns of content sharing essentially harvests the collective intelligence of sources on the social medium. Individuals who receive the shared content interpret it in proper context then act (e.g., forward or not) in accordance with their biases, generating a distinct bias-specific propagation pattern that the automated separation algorithm captures. 

The aforementioned approach for content separation by propagation pattern also allows automated identification of {\em neutrally-worded\/} posts (and their sources), as those posts tend to propagate {\em in both bias groups\/}. Content that does not propagate strictly on group boundaries is likely to be less biased. The insight leads to solutions for understanding {\em overlap\/} in beliefs~\cite{yang2020disentangling}. In turn, these solutions may help with automated de-biasing of collected information, possibly under the control of the end user. Metrics can also be designed to help users understand the degree of bias in content they are exposed to. Prior work has shown that excluding posts that fail the de-biasing test tends to increase the quality of remaining content~\cite{al2014crowd}. 
In general, the approach described above allows developing services that give consumers control over the degree of bias in information they ingest. For example, they may choose to sample both sides, lean on one side, or choose unbiased content. Such explicit control  does not exist with current curation services. This control might be most effective when combined with the technique described next; namely, information summarization. 



%% file: 04-HengGeneration.tex
\section{Mitigating Information Overload}
Information overload and the resulting need for content filtering is the main contributor to selective exposure. Figure~\ref{fig:load} suggests that if content coverage (by a user) was increased, the extent of fragmentation will be reduced. How can one achieve this effect?

Human cognitive capacity cannot be suddenly increased (without a leap in evolution). In the meantime, accessible information is likely to continue to grow at an accelerated rate. The approach taken by current curation services (e.g., search and ranking engines) relies on {\em extractive\/} summarization. Namely, it picks {\em individual information items\/} that best match the search query. This technique is fundamentally unsustainable in view of the continued growth in information volume; the ``best matches'' will constitute progressively smaller fractions of relevant content. 
We believe the next leap in curation will therefore need to adopt {\em abstractive\/} summarization: a distillation of content that directly conveys the main points. Abstractive summarization increases the effective fraction of content that an individual can digest within their capacity limits. It therefore reduces the impact of information overload (which, as Figure~\ref{fig:load} shows, ameliorates ideological fragmentation).

How does one summarize the past events? 
Our minds represent events at various levels of granularity and abstraction, which allows us to quickly access and reason about old and new scenarios. Progress in natural language understanding and computer vision has helped automate some parts of event understanding. Still, the current, first-generation, automated event understanding is overly simplistic since it is local, sequential and flat. Real events are hierarchical and probabilistic. Understanding them requires knowledge in the form of a repository of abstract event schemas (complex event templates), understanding the progress of time, using background knowledge, and performing global inference. When complex events unfold in an emergent and dynamic manner, the multimedia multilingual digital data from traditional news media and social media often convey conflicting information. To understand the many facets of such complex, dynamic situations, one needs cross-media cross-document event coreference resolution and event-event relation tracking methods to gain event-centric knowledge of such situations. 
These techniques can leverage solutions for algorithmic misinformation detection and bias mitigation, described earlier, to offer more reliable versions of transpired events. 
Current advances on this topic are overviewed in a recent survey of abstractive summarization techniques~\cite{moratanch2016survey}.

%% file: 05-Dialogue.tex
\section{Restoring Dialogue}
The above three solutions are geared at counteracting the three influences that feed the fragmentation cycle, shown in Figure~\ref{fig:cycle}. However, they do not have a {\em restorative\/} effect. Detecting bad sources, reducing bias, and correctly summarizing events is not the end of polarizing conflicts. Can computing technology also help de-escalate ideological confrontation?

Many conflicts arise due to differences in moral choices. For example, in a local election (taking inspiration from the COVID 19 crisis in 2020), some narratives might support a governor's stance on upholding community interests (such as curbing the spread of an infectious disease by ordering citizens to shelter in place), whereas others might bring up the toll it took on individuals who consequently lost their livelihoods. This is a moral conflict between individual freedoms and interests of the collective.

Moral Foundations Theory~\cite{graham2013moral}  decomposes moral values into five distinct categories, each defined by its virtue/vice pair. These categories are (i) {\em care/harm\/}, (ii) {\em fairness/discrimination\/}, (iii) {\em loyalty/betrayal\/}, (iv) {\em authority/subversion\/}, and (v) {\em purity/degradation\/}~\cite{graham2013moral}. While evolution (according to this theory) predisposed us to adopt behaviors that align with these values (because they offered an evolutionary advantage), the theory suggests that different cultural contexts reinforce certain values more than others, leading to a divergence in moral judgment across communities. By understanding the values reinforced by a particular group, it is possible to better understand their position on issues. 
Understanding moral foundations may in principle help dialogue across groups. An argument presented from a moral perspective that agrees with the recipient's beliefs may resonate better with the recipient.  Such an understanding can be {\em automated\/} from analysis of arguments presented by the group. Specifically, each moral foundation is associated with a vocabulary of words that are more likely to be used when arguing from that perspective~\cite{sagi2014measuring}. For example, looking at Table~\ref{eurovision_showcase}, it can be seen that Jamala supporters argue for Jamala on the basis of care/harm (with words like ``suffering'', ``abuse'', ``harrassment'' and ``trolls'' suggesting compassion with her position). On the other hand, opposing voices argue primarily based on authority/subversion, mentioning ``violation'' of ``rules''. 

Understanding a group's moral perspective may help craft better arguments when presenting a given position for that group. A prominent example of argument customization to different moral foundations can be found in excerpts from a speech by Barack Obama (the US president at the time) to rally US support in 2014 for action against ISIL, the self-proclaimed ``Islamic State'' in Iraq and the Levant (also known as ISIS).\footnote{Full speech available at: https://time.com/3320666/obama-isis-speech-full-text/} In his speech, he stated that: ``ISIL is not Islamic'', ``ISIL is certainly not a state'' (an {\em authority\/} argument that asserts the lack of a legitimate authority to protect ISIL). The speech continued: ``it has no vision other than the slaughter of [innocents]'' (a {\em care/harm\/} argument suggesting that they inflict indiscriminate harm). He also asserted that: ``They enslave, rape, and force women into marriage'', ``They threatened a religious minority with genocide'' (a {\em fairness/discrimination\/} argument describing gender-based and religious discrimination). 
Finally, he observed that: ``They took the lives of two American journalists'' (a {\em loyalty\/} argument suggesting that it is patriotic of his nation to stand together against this threat). 

Can future linguistic and machine intelligence tools automatically generate convincing arguments from a specified moral perspective? IBM's Project Debater\footnote{https://www.research.ibm.com/artificial-intelligence/project-debater/live/} aims to develop the first machine-learning system that automates the construction of arguments to support a position. Can a similar technique be used as part of information summarization to capture to an individual the positions taken by different parties and the moral arguments behind them?